\documentclass{llncs}
\usepackage{llncsdoc}
\usepackage{booktabs}
\usepackage{slashbox}
\usepackage{url}
\usepackage{amssymb}
\usepackage{amsmath}
\usepackage{wrapfig}
\usepackage{enumitem}
\usepackage{mdwlist}
\usepackage[caption=false]{subfig}
\usepackage{microtype}
\usepackage{color}
\usepackage{capt-of}
\usepackage{eurosym}
\usepackage{algorithm}
\usepackage{array}
\usepackage[pdftex]{graphicx} 
\usepackage{longtable}
\newcolumntype{L}[1]{>{\raggedright\let\newline\\\arraybackslash\hspace{0pt}}m{#1}}
\newcolumntype{C}[1]{>{\centering\let\newline\\\arraybackslash\hspace{0pt}}m{#1}}
\newcolumntype{R}[1]{>{\raggedleft\let\newline\\\arraybackslash\hspace{0pt}}m{#1}}
\usepackage{rotating}
\usepackage{makecell}
\RequirePackage[normalem]{ulem} 
\RequirePackage{color}\definecolor{RED}{rgb}{1,0,0}\definecolor{BLUE}{rgb}{0,0,1} 
\usepackage{tikz}
\usetikzlibrary{arrows,automata,backgrounds,decorations,fit,petri,positioning,petri,shapes,calc,patterns,shadows}
\tikzset{small/.style={level distance=10pt,sibling distance=-2pt,outer ysep=-2pt, label distance=-8pt}}
\tikzset{smaller/.style={level distance=20pt}}
\tikzset{smallish/.style={level distance=20pt,sibling distance=-2pt,outer ysep=-2pt, label distance=-8pt}}
\tikzset{node distance=5mm} 
\tikzset{place/.append style={circle,draw=black,thick,inner sep=0pt,minimum size=4mm,label position=below}} 
\tikzset{transition/.append style={rectangle,draw=black,thick,inner sep=0pt,minimum width=24mm, minimum height=8mm}}
\tikzset{tau/.style={transition,fill=black}}  

\newcommand{\fleche}{\longrightarrow}
\newcommand{\flsup}[1]{\stackrel{#1}{\fleche}}
\newcommand{\step}[1]{\flsup{#1}}           
\newcommand{\Lan}                 {\mathfrak{L}}

\newcommand{\pre}[1]{\bullet #1}
\newcommand{\APN}{\mathit{APN}}

\newcommand{\MF}{\mathit{MF}}

\newlength{\hatchspread}
\newlength{\hatchthickness}
\newlength{\hatchshift}
\newcommand{\hatchcolor}{}
\tikzset{hatchspread/.code={\setlength{\hatchspread}{#1}},
	hatchthickness/.code={\setlength{\hatchthickness}{#1}},
	hatchshift/.code={\setlength{\hatchshift}{#1}},
	hatchcolor/.code={\renewcommand{\hatchcolor}{#1}}}
\tikzset{hatchspread=3pt,
	hatchthickness=0.4pt,
	hatchshift=0pt,
	hatchcolor=black}
\pgfdeclarepatternformonly[\hatchspread,\hatchthickness,\hatchshift,\hatchcolor]
{custom north west lines}
{\pgfqpoint{\dimexpr-2\hatchthickness}{\dimexpr-2\hatchthickness}}
{\pgfqpoint{\dimexpr\hatchspread+2\hatchthickness}{\dimexpr\hatchspread+2\hatchthickness}}
{\pgfqpoint{\dimexpr\hatchspread}{\dimexpr\hatchspread}}
{
	\pgfsetlinewidth{\hatchthickness}
	\pgfpathmoveto{\pgfqpoint{0pt}{\dimexpr\hatchspread+\hatchshift}}
	\pgfpathlineto{\pgfqpoint{\dimexpr\hatchspread+0.15pt+\hatchshift}{-0.15pt}}
	\ifdim \hatchshift > 0pt
	\pgfpathmoveto{\pgfqpoint{0pt}{\hatchshift}}
	\pgfpathlineto{\pgfqpoint{\dimexpr0.15pt+\hatchshift}{-0.15pt}}
	\fi
	\pgfsetstrokecolor{\hatchcolor}
	\pgfusepath{stroke}
}

\begin{document}
\title{Interest-Driven Discovery of Local Process Models}

\urldef{\mailsa}\path|{n.tax,n.sidorova,w.m.p.v.d.aalst}@tue.nl|
\urldef{\mailsb}\path|{benjamin.dalmas,sylvie.norre}@isima.fr|

\author{Niek Tax\inst{1} \and Benjamin Dalmas\inst{2} \and Natalia Sidorova\inst{1} \and Wil M.P. van der Aalst\inst{1}\and Sylvie Norre\inst{2}}

\institute{Eindhoven University of Technology, Department of Mathematics and Computer Science, P.O. Box 513, 5600MB Eindhoven, The Netherlands\\
\mailsa
\and Clermont-Auvergne University, LIMOS CNRS UMR 6158, Aubi\`{e}re, France\\
\mailsb
}

\maketitle
\setcounter{footnote}{0}

\begin{abstract}
Local Process Models (LPM) describe structured fragments of process behavior occurring in the context of less structured business processes. Traditional LPM discovery aims to generate a collection of process models that describe highly frequent behavior, but these models do not always provide useful answers for questions posed by process analysts aiming at business process improvement. We propose a framework for \emph{goal-driven LPM discovery}, based on utility functions and constraints. We describe four scopes on which these utility functions and constrains can be defined, and show that utility functions and constraints on different scopes can be combined to form composite utility functions/constraints. Finally, we demonstrate the applicability of our approach by presenting several actionable business insights discovered with LPM discovery on two real life data sets.
\end{abstract}


\section{Introduction}





Process Mining \cite{Aalst2016} has emerged as a new discipline aiming at the improvement of business processes through the analysis of event data recorded by information systems. Such event logs capture the different steps (events) that are recorded for each instance of the process (case), and record for each of those steps what was done, by whom, for whom, where, when, etc. Process discovery, one of the main tasks in the process mining field, is concerned with the discovery of an interpretable model from this event log such that this model accurately describes the process. The process models obtained give insight in what is happening in the process, and can be used as a starting point for different types of further analysis, e.g. bottleneck analysis \cite{Maruster2009}, and checking compliance with rules and regulations \cite{Ramezani2012}. Many algorithms have been proposed for process discovery, e.g., \cite{Bergenthum2007,Maggi2011,Leemans2013,Liesaputra2015,Gunther2007} (see Section \ref{sec:related_work}).\looseness=-1
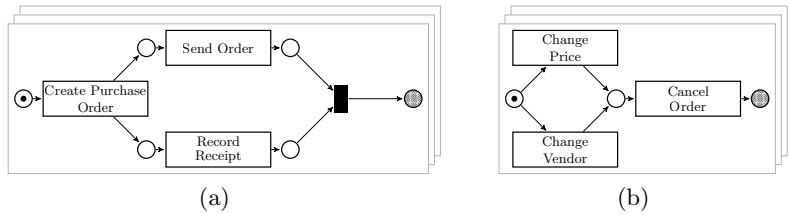
\begin{figure}[t]
	\centering
	\subfloat[]{
		\scalebox{0.58}{
			\begin{tikzpicture}
			[node distance=1.65cm,
			on grid,>=stealth',
			bend angle=20,
			auto,
			every place/.style= {minimum size=5mm},
			transitionH/.style={rectangle, thick, fill=black, minimum width=3mm, inner ysep=9pt },
			framed,
			scale=3, 
			blend group=screen,
			background rectangle/.style={
				double copy shadow={shadow xshift=1ex,shadow
					yshift=+1.5ex}, fill=white, draw=black!30}
			]
			\node [place, tokens = 1] (p){};
			\node [transition] (2) [align=center, right = of p] {Create Purchase\\Order}
			edge [pre] node[auto] {} (p);
			\node [place] (p3) [above right = of 2] {}
			edge[pre] node[auto] {} (2);
			\node [transition] (t1) [right = of p3]{Send Order}
			edge[pre] node[auto] {} (p3);
			\node [place] (p4) [below right = of 2] {}
			edge[pre] node[auto] {} (2);
			\node [transition] (t2) [right = of p4] {\shortstack{Record\\Receipt}}
			edge[pre] node[auto] {} (p4);
			\node [place] (p5) [right = of t1] {}
			edge[pre] node[auto] {} (t1);
			\node [place] (p6) [right = of t2] {}
			edge[pre] node[auto] {} (t2);
			\node [transitionH] (t3) [below right = of p5] {}
			edge[pre] node[auto] {} (p5)
			edge[pre] node[auto] {} (p6);
			\node [place,pattern=custom north west lines,hatchspread=1.5pt,hatchthickness=0.25pt,hatchcolor=gray] (p7) [right=of t3] {}
			edge[pre] node[auto] {}(t3);
			\end{tikzpicture}}
		\label{sfig:motivating_example_1}	
	}
	\qquad
	\subfloat[]{
		\scalebox{0.58}{
			\begin{tikzpicture}
			[node distance=1.65cm,
			on grid,>=stealth',
			bend angle=20,
			auto,
			every place/.style= {minimum size=5mm},
			transitionH/.style={rectangle, thick, fill=black, minimum width=3mm, inner ysep=9pt },
			framed,
			scale=3, 
			blend group=screen,
			background rectangle/.style={
				double copy shadow={shadow xshift=1ex,shadow
					yshift=+1.5ex}, fill=white, draw=black!30}
			]
			\node [place, tokens = 1] (p){};
			\node [transition] (t1) [above right = of p]{\shortstack{Change\\Price}}
			edge[pre] node[auto] {} (p);
			\node [transition] (t2) [below right = of p] {\shortstack{Change\\Vendor}}
			edge[pre] node[auto] {} (p);
			\node [place] (p4) [below right = of t1] {}
			edge[pre] node[auto] {} (t1)
			edge[pre] node[auto] {} (t2);
			\node [transition] (2) [align=center, right = of p4] {\shortstack{Cancel\\ Order}}
			edge [pre] node[auto] {} (p4);
			\node [place,pattern=custom north west lines,hatchspread=1.5pt,hatchthickness=0.25pt,hatchcolor=gray] (p7) [right=of 2] {}
			edge[pre] node[auto] {}(2);
			\end{tikzpicture}}
		\label{sfig:motivating_example_2}
	}
	\caption{\emph{(a)} A frequent Local Process Model with low utility, and \emph{(b)} a non-frequent Local Process Model with high utility.}
	\label{fig:motivating_example}
	\vspace{-0.6cm}
\end{figure}

One type of process discovery is Local Process Model (LPM) discovery \cite{Tax2016,Tax2016b}, which is concerned with the discovery of a ranking of process models, where each individual LPM describes only a subset of the process activities. Each LPM describes one frequent pattern in a process model notation (e.g. Petri net \cite{Murata1989}, BPMN \cite{OMG2011}, or UML activity diagram \cite{ISO2012}). This gives an LPM the full expressive power of the respective process model notation and allows it to represent more complex non-binary relations that cannot be expressed in declarative process models. LPMs aim to describe frequent local pieces of behavior, therefore, LPMs can be seen as a special form of \emph{frequent pattern mining} \cite{Han2007} where each pattern is a process model. However, LPMs are not limited to subsequences \cite{Srikant1996} or episodes \cite{Leemans2014}.\looseness=-1

A recent trend in the frequent pattern mining field is to incorporate \emph{utility} into the pattern selection framework, such that not just the most frequent patterns are discovered, but instead patterns are discovered that address typical business concerns, such as the patterns that represent high financial costs. Shen et al. \cite{Shen2003} were the first to introduce utility-based itemset mining. Since then, utility-based pattern mining has spread to different types of pattern mining, including sequential pattern mining \cite{Yin2012}. Utility-based pattern mining techniques assume that the value of each data point with respect to a certain business question is known and then discover the optimal patterns in terms of the value that they represent.\looseness=-1 

Imagine a Purchase-to-Pay process and as a process analyst we are interested in where in the process the employees of the company spend most of their time. Figure \ref{fig:motivating_example} shows two LPMs that could be discovered from such an event log. Figure \ref{sfig:motivating_example_1} is a process fragment that describes the creation of a purchase order, which is followed by both the sending of the order and the recording of the receipt in an arbitrary order. This process fragment is frequent, as they are required steps for each order. Figure \ref{sfig:motivating_example_2} describes a process fragment where the order is canceled after the price or the vendor of the order is changed. Even though the process fragment of Figure \ref{sfig:motivating_example_2} is likely to be infrequent, it will take considerable resources of the department as canceling an order is an undesired action and considerable time will be spend trying to prevent it. Existing support-based LPM discovery \cite{Tax2016} would not be able to discover Figure \ref{sfig:motivating_example_2} because of its low frequency, motivating the need for utility-based LPM discovery.\looseness=-1

In this paper we propose a framework to discover LPMs based on their utility in the context of a particular business question. Furthermore, we give an extensive overview of utility functions and their relevance in a BPM context. The techniques described in this paper have been implemented in the \emph{ProM} process mining framework \cite{Dongen2005} as part of the \emph{LocalProcessModelDiscovery}\footnote{\url{https://svn.win.tue.nl/trac/prom/browser/Packages/LocalProcessModelDiscovery}} package.\looseness=-1



This paper is organized as follows. Section \ref{sec:related_work} describes related work. Section \ref{sec:preliminaries} introduces the basic concepts used in this paper. Section \ref{sec:utility_functions} introduces utility functions and constraints in the context of LPMs. In Section \ref{sec:case_study_1} we demonstrate utility-based LPM discovery on two real-life event logs and show that we can obtain actionable insights. 
We conclude the paper in Section \ref{sec:conclusion}.
\section{Related Work}
\label{sec:related_work}
In this section we discuss two areas of related work. First we discuss existing work in process discovery and position Local Process Model (LPM) discovery in the process discovery landscape. Secondly, we discuss related work from the pattern mining field.
\subsection{Process Discovery}
Process discovery techniques can be classified in several dimensions. Some process discovery techniques discover \emph{formal process models}, where the behavior allowed by the model is formally defined, while others discover  \emph{informal process models} with unclear semantics. Orthogonally, process discovery algorithms can be classified in \emph{end-to-end (global)} techniques that produce models that describe the logged process executions fully from start to end, and \emph{pattern-based (local)} techniques that produce models that describe the behavior of the log only partially. Table \ref{tab:process_discovery} provides a classification of some existing process discovery techniques, and shows that LPM discovery is the only technique available which provides models that are both formal and local.
\begin{table}[t]
\centering
\scalebox{0.8}{
\begin{tabular}{|l|c|c|}
	\toprule
	Algorithm & Formal/Informal & Global/Local \\
	\midrule
	Declare Miner\cite{Maggi2011} & Formal & Global\\
	Language-based regions \cite{Bergenthum2007} & Formal & Global\\
	Inductive Miner \cite{Leemans2013} & Formal & Global\\
	LPM Discovery \cite{Tax2016} & Formal & Local\\
	Fuzzy Miner \cite{Gunther2007} & Informal & Global\\
	Episode Miner \cite{Leemans2014} & Informal & Local\\
	\bottomrule
\end{tabular}}
\caption{A classification of process discovery methods.}
\label{tab:process_discovery}
\vspace{-0.7cm}
\end{table}

\subsection{Interest-driven Pattern Mining}
LPMs are able to express a richer set of relations (e.g. loops, XOR-constructs, concurrency) than sequential pattern mining techniques, which are limited to the sequential constructs. The mining of patterns driven by the interest of an analyst is known in the pattern mining field as high-utility pattern mining. Several high-utility sequential pattern mining algorithms have been proposed \cite{Yin2012,Lan2014,Yin2013}. Sequential pattern mining techniques take as input a \emph{sequence database}, a concept which is closely related to \emph{event logs} in process mining. Sequence databases consist of sequences of tasks (called \emph{items}). Some pattern mining techniques additionally assume the timestamp of each item to be logged. Sequence databases are not as rich as events logs found in process mining, where many data attributes are logged both for events and cases.

A second approach to interest-driven pattern mining pattern is constraint-based sequential pattern mining. In contrast to high-utility pattern mining, which gives preference to more useful patterns, constraint-based pattern mining completely removes non-useful ones. Pei et al. \cite{Pei2007} provides a categorization of pattern constraints, consisting of four types of constraints on the ordering of items in the pattern and two types of constraints on the timestamps of the items in the log that are instances of a pattern. This richer notion of an event log allows us to define utility and constraints in a more flexible way (e.g. using event or case properties) compared to high-utility pattern mining techniques which define utility as mapping from an item type to utility and constraint-based pattern mining which limit the utility definition to ordering information and the time domain.\looseness=-1
\section{Basic Definitions}
\label{sec:preliminaries}
In this section we introduce notations related to event logs, Petri nets, and Local Process Models (LPMs), which are used in later sections of this paper.

\subsection{Events, Traces, and Event Logs}

For a given set $A$, $A^*$ denotes the set of all sequences over $A$ and $\sigma=\langle a_1,a_2,\dots,a_n\rangle$ is a sequence of length $n$, where $\sigma(i)=a_i$. $\langle\rangle$ is the empty sequence and $\sigma_1\sigma_2$ is the concatenation of sequences $\sigma_1$ and $\sigma_2$. $\sigma{\upharpoonright}_{A}$ is the projection of $\sigma$ on $A$, e.g. $\langle a,b,c,a,b,c\rangle{\upharpoonright}_{\{a,c\}}{=}\langle a,c,a,c \rangle$.

Let $\mathcal{E}$ be the event universe, i.e., the set of all possible event identifiers. We assume that events are characterized by various properties, e.g., an event has a timestamp, corresponds to an activity, is performed by a particular resource, etc. We do not impose a specific set of properties, however, we assume that two of these properties are the activity and timestamp of an event, i.e., there is a function $\pi_\mathit{activity}: \mathcal{E}\rightarrow\mathcal{A}$ that assigns to each event an activity from a finite set of process activities $\mathcal{A}$, and a function $\pi_\mathit{time}: \mathcal{E}\rightarrow\mathcal{T}$ that maps each event to the time domain $\mathcal{T}$. In general, we write $\pi_\mathit{p}(e)$ to obtain the value of any property $p$ of event $e$.\looseness=-1


An \emph{event log} is a set of events, each linked to one trace and globally unique, i.e., the same event cannot occur twice in a log. A trace in a log represents the execution of one case. A \emph{trace} is a finite non-empty sequence of events $\sigma{\in}\mathcal{E}^*$ such that each event appears only once and time is non-decreasing, i.e., for $1{\le}i{<}j{\le}|\sigma|:\sigma(i){\neq}\sigma(j)$ and $\pi_\mathit{time}(\sigma(i)){\le}\pi_\mathit{time}(\sigma(j))$. $\mathcal{C}$ is the set of all possible traces. An \emph{event log} is a set of traces $L{\subseteq}\mathcal{C}$ such that each event appears at most once in the entire log.

Given a trace and a property, we often need to compute a sequence consisting of the value of this property for each event in the trace. To this end, we lift the function $\pi_p$ that maps an event to the value of its property $p$, in such a way that we can apply it to sequences of events (traces).

A partial function $f \in A \nrightarrow Y$ with domain $dom(f)$ can be lifted to sequences over $A$ using the following recursive definition: (1) $f(\langle\rangle)=\langle\rangle$;  (2) for any $\sigma\in A^*$ and $x\in A$:
\begin{center}
	$f(\sigma \cdot \langle x\rangle) =
	\left\{
	\begin{array}{ll}
	f(\sigma)  & \mbox{if } x{\notin}\mathit{dom}(f), \\
	f(\sigma) \cdot \langle f(x)\rangle & \mbox{if } x{\in}\mathit{dom}(f).
	\end{array}
	\right.$
\end{center}
$\pi_\mathit{activity}(\sigma)$ transforms a trace $\sigma$ to a sequence of its activities. For example, for trace $\sigma{=}\langle e_1, e_2\rangle$ with $\pi_\mathit{activity}(e_1){=}a$ and $\pi_\mathit{activity}(e_2){=}b$: $\pi_\mathit{activity}(\sigma){=}\langle a,b\rangle$.

Traces themselves can also have properties, e.g., a case represented by trace $\sigma{\in}L$ can be associated with a branch of the company where the process was executed. We write $\phi_p(\sigma)$ to obtain the value of any property $p$ of a case represented by trace $\sigma$.

\begin{figure}[t]
	\centering
	\includegraphics[width=0.8\linewidth]{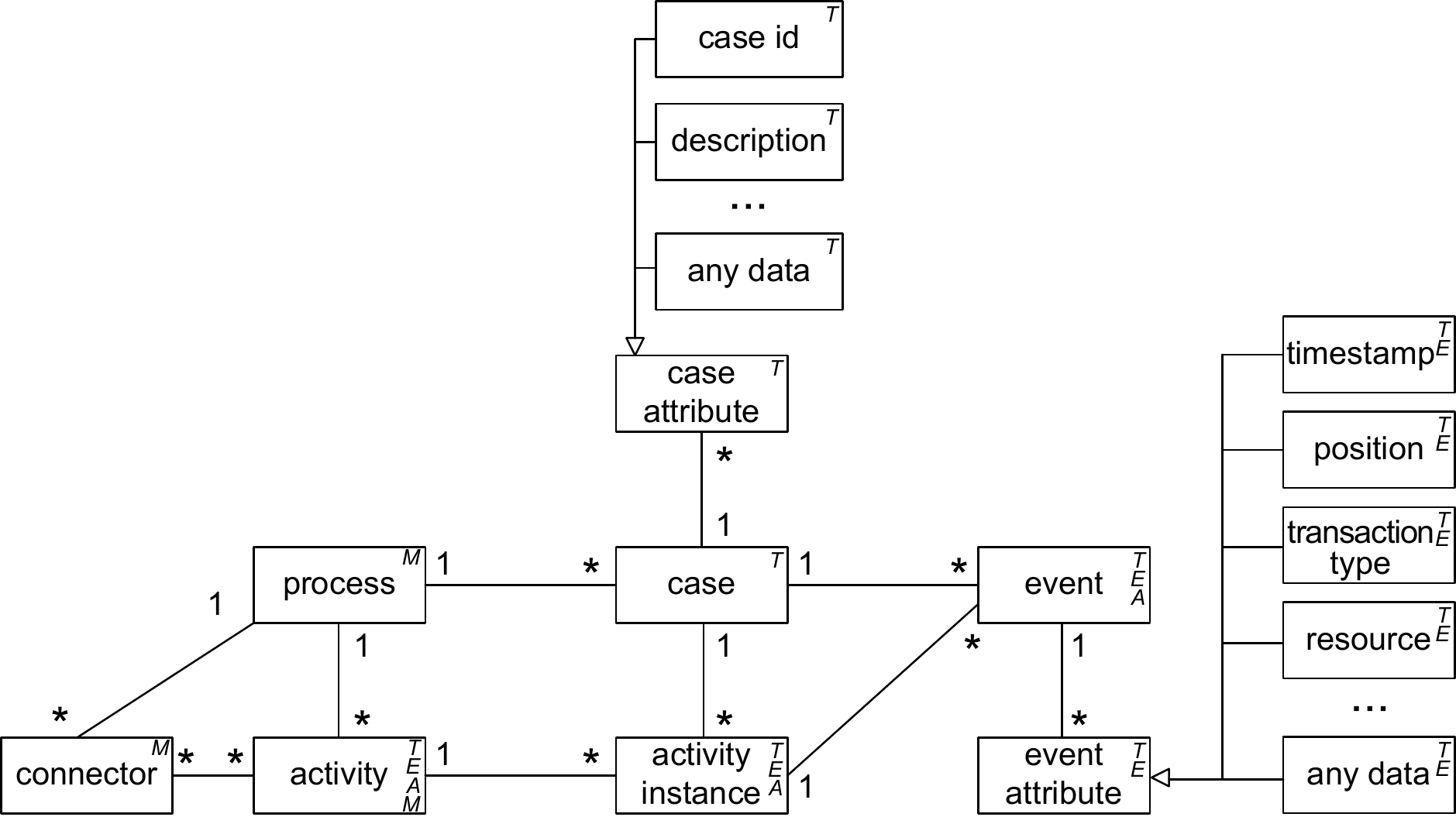}
	\caption{Basic logging concepts conceptualized in a class diagram.}
	\label{fig:scope}
	\vspace{-0.4cm}
\end{figure}

The class diagram in Figure \ref{fig:scope} is obtained from \cite{Aalst2016} and it conceptualizes basic logging concepts in process mining. For now, ignore the annotations on the right side of the classes (\emph{T,E,A,M}). A process model consists of a set of process activities and a \emph{connector} connects elements of the process models to those activities.\looseness=-1 
\subsection{Petri nets}
A process model notation that is frequently used in the process mining area is the Petri net.
\begin{definition}[Labeled Petri net]
	\label{def:lpn}
	A \emph{labeled Petri net} $N=\langle P,T,F,\Sigma_M,\ell\rangle$ is a tuple where $P$ is a finite set of places, $T$ is a finite set of transitions such that $P \cap T = \emptyset$,  $F \subseteq (P \times T) \cup (T \times P)$ is a set of directed arcs, called the flow relation, $\Sigma_M$ is a finite set of labels representing activities, and $\ell:T\nrightarrow \Sigma_M$ is a labeling function that assigns a label to transitions. Unlabeled transitions, i.e., $t{\in}T$ with $t{\not}{\in}dom(l)$, are referred to as $\tau$-transitions, or invisible transitions.
\end{definition}
For a node $n\in P{\cup}T$ we use $\bullet n$ and $n \bullet$ to denote the set of input and output nodes of $n$, defined as $\bullet n =\{n'|(n',n){\in}F\}$ and $n \bullet =\{n'|(n,n'){\in}F\}$ . 

A state of a Petri net is defined by its \emph{marking} $M{\in}\mathbb{N}^{P}$ being a multiset of places. A marking is graphically denoted by putting $M(p)$ tokens on each place $p\in P$. State changes occur through transition firings. A transition $t$ is enabled (can fire) in a given marking $M$ if each input place $p\in \pre{t}$ contains at least one token. Once a transition fires, one token is removed from each input place of $t$  and one token is added to each output place of $t$, leading to a new marking $M'$ defined as $M'=M-\bullet t+t\bullet$.
A firing of a transition $t$ leading from marking $M$ to marking $M'$ is denoted as $M \step{t} M'$. $M_1 \step{\sigma} M_2$ indicates that $M_2$ can be reached from $M_1$ through a firing sequence $\sigma\in T^*$.\looseness=-1

Often it is useful to consider a Petri net in combination with an initial marking and a set of possible final markings. This allows us to define the language accepted by the Petri net and to check whether some  behavior is part of the behavior of the Petri net (can be replayed on it).

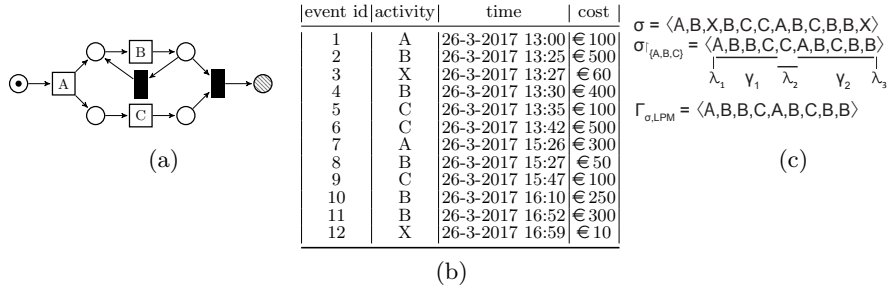
\begin{figure}[t]
	\centering
	\captionsetup[subfigure]{width=4.5cm}
	\subfloat[]{
	\scalebox{0.6}{
		\begin{tikzpicture}
		[node distance=1cm,
		on grid,>=stealth',
		bend angle=20,
		auto,
		every place/.style= {minimum size=5mm},
		transition/.append style={rectangle,draw=black,thick,inner sep=0pt,minimum size=5mm},
		transitionH/.style={rectangle, thick, fill=black, minimum width=3mm, inner ysep=9pt }
		]
		\node [place, tokens = 1] (p){};
		\node [transition] (2) [align=center, right = of p] {A}
		edge [pre] node[auto] {} (p);
		\node [place] (p3) [above right = of 2] {}
		edge[pre] node[auto] {} (2);
		\node [transition] (t1) [right = of p3]{B}
		edge[pre] node[auto] {} (p3);
		\node [place] (p4) [below right = of 2] {}
		edge[pre] node[auto] {} (2);
		\node [transition] (t2) [right = of p4] {C}
		edge[pre] node[auto] {} (p4);
		\node [place] (p5) [right = of t1] {}
		edge[pre] node[auto] {} (t1);
		\node [transitionH] (t4) [below = 0.72 of t1,label=right:{}] {}
		edge[pre] node[auto] {} (p5)
		edge[post] node[auto] {} (p3);
		\node [place] (p6) [right = of t2] {}
		edge[pre] node[auto] {} (t2);
		\node [transitionH] (t3) [below right = of p5,label=below right:{}] {}
		edge[pre] node[auto] {} (p5)
		edge[pre] node[auto] {} (p6);
		\node [place,pattern=custom north west lines,hatchspread=1.5pt,hatchthickness=0.25pt,hatchcolor=gray] (p7) [right=of t3] {}
		edge[pre] node[auto] {}(t3);
		\end{tikzpicture}}
	\label{sfig:example_apn}	
	}
	\subfloat[]{
	\scalebox{0.75}{
	\def\arraystretch{0.8}
	\begin{tabular}{|c|c|c|c|}
		\toprule
		event id&activity&time&cost\\
		\midrule
		1 & A & 26-3-2017 13:00 & \EUR{100}\\
		2 & B & 26-3-2017 13:25 & \EUR{500}\\
		3 & X & 26-3-2017 13:27 & \EUR{60}\\
		4 & B & 26-3-2017 13:30 & \EUR{400}\\
		5 & C & 26-3-2017 13:35 & \EUR{100}\\
		6 & C & 26-3-2017 13:42 & \EUR{500}\\
		7 & A & 26-3-2017 15:26 & \EUR{300}\\
		8 & B & 26-3-2017 15:27 & \EUR{50}\\
		9 & C & 26-3-2017 15:47 & \EUR{100}\\
		10 & B & 26-3-2017 16:10 & \EUR{250}\\
		11 & B & 26-3-2017 16:52 & \EUR{300}\\
		12 & X & 26-3-2017 16:59 & \EUR{10}\\
		\bottomrule
	\end{tabular}
	\label{sfig:example_trace}
	}}
	\subfloat[]{
	\centering
	\includegraphics[width=0.28\linewidth]{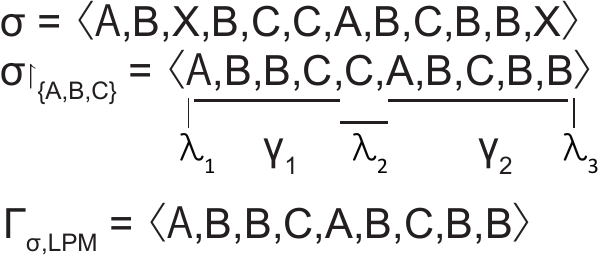}
	\label{sfig:example_alignment}
	\vspace{-0.3cm}
}
	\caption{\emph{(a)} Example accepting Petri net $APN_1$. \emph{(b)} Trace $\sigma$ of an event log $L$. \emph{(c)} Segmentation of $\sigma$ on $APN_1$.}
	\label{fig:example}
	\vspace{-0.5cm}
\end{figure}
\begin{definition}[Accepting Petri net]
	An \emph{accepting Petri net} is a triple $\APN{=}(N,M_0,\MF)$, where $N$ is a labeled Petri net, $M_0\in\mathbb{N}^P$ is its initial marking, and $\MF\subseteq\mathbb{N}^P$ is its set of possible final markings, such that $\forall_{M_1,M_2 \in \MF} M_1{\nsubseteq}M_2$. A sequence $\sigma{\in}T^*$ is called a \emph{trace} of an accepting Petri net $\APN$ if $M_0\step{\sigma} M_f$ for some final marking $M_f{\in}\MF$. The \emph{language} $\Lan(\mathit{APN})$ of $\APN$ is the set of all label sequences belonging to its traces, i.e., if $\sigma$ is a trace, then $l(\sigma){\in}\Lan(\mathit{APN})$.
\end{definition}
Figure \ref{sfig:example_apn} shows an example of an accepting Petri net. Circles represent places and rectangles represent transitions. Invisible transitions ($\tau$) are depicted as black rectangles. Places that belong to the initial marking contain a token and places belonging to a final marking contain a bottom right label $f_i$ with $i$ a final marking identifier, or are simply marked as $\begin{tikzpicture}
[node distance=1.4cm,
on grid,>=stealth',
bend angle=20,
auto,
every place/.style= {minimum size=0.1mm},
]
\node [place,pattern=custom north west lines,hatchspread=1.5pt,hatchthickness=0.25pt,hatchcolor=gray] {};
\end{tikzpicture}$ in case of a single final marking. The language of this accepting Petri net, with $\Sigma_M=\{A,B,C\}$ is $\{\langle A,B,C\rangle,\langle A,C,B\rangle,\langle A,B,B,C\rangle,\langle A,B,C,B\rangle,\langle A,B,B,B,C\rangle,\dots\}$. We refer the interested reader to \cite{Murata1989} for a more thorough introduction of Petri nets. 

\subsection{Local Process Models}
LPMs \cite{Tax2016} are process models that describe the behavior seen in the event log only partially, focusing on frequently observed behavior. Typically, LPMs describe the behavior of only up to 5 activities. LPMs can be represented in any process modeling notation, such as BPMN \cite{OMG2011}, UML \cite{ISO2012}, or EPC \cite{Keller1992}. Here we use Petri nets to represents LPMs. A technique to generate a ranked collection of LPMs through iterative expansion of candidate process models is proposed in \cite{Tax2016}. The search space of process models is fixed, depending on the event log. We define $\mathit{LPMS}(L)$ as the set of possible LPMs that can be constructed for given event log $L$. We refer the reader to \cite{Tax2016} for a detailed description of search space $\mathit{LPMS}(L)$.\looseness=-1

To evaluate a given LPM on a given event log $L$, its traces $\sigma{\in}L$ are first projected on the set of activities $\Sigma_M$ in the LPM, i.e., $\sigma'{=}\sigma{\upharpoonright}_{\Sigma_M}$. The projected trace $\sigma'$ is then segmented into $\gamma$-segments that fit the behavior of the LPM and $\lambda$-segments that do not fit the behavior of the LPM, i.e., $\sigma'{=}\lambda_1 \gamma_1 \lambda_2 \gamma_2 \cdots \lambda_n \gamma_n \lambda_{n+1}$ such that $\gamma_i{\in}\Lan(\mathit{LPM})$ and $\lambda_i{\not\in}\Lan(\mathit{LPM})$. We define $\Gamma_{\sigma,LPM}$ to be a function that projects trace $\sigma$ on the LPM activities and obtains its subsequences that fit the LPM, i.e., $\Gamma_{\sigma,LPM}=\gamma_1\gamma_2\dots\gamma_n$.

Let our LPM under evaluation be the Petri net of Figure \ref{sfig:example_apn} and let Figure \ref{sfig:example_trace} be our trace $\sigma$, with $\pi_\mathit{activity}(\sigma)=\langle A,B,X,B,C,C,A,B,C,B,B,X\rangle$. Projection on the activities of the LPM gives $\pi_\mathit{activity}(\sigma){\upharpoonright}_{\{A,B,C\}}=\langle A,B,B,C,C,\\A,B,C,B,B\rangle$. Figure \ref{sfig:example_alignment} shows the segmentation the projected traces on the LPM, leading to $\pi_\mathit{activity}(\Gamma_{\sigma,LPM})=\langle A,B,B,C,A,B,C,B,B\rangle$. The segmentation starts with an empty non-fitting segment $\lambda_1$, followed by fitting segment $\gamma_1{=}\langle A,B,B,C\rangle$, which completes one run through the model from initial to final marking. The second event $C$ in $\sigma$ cannot be replayed on $LPM$, since it only allows for one $C$ and $\gamma_1$ already contains a $C$. This results in a non-fitting segment $\lambda_2{=}\langle C\rangle$. $\gamma_2{=}\langle A,B,C,B,B\rangle$ again represents a run through the model from initial to final marking, and $\lambda_3{=}\langle D\rangle$ does not fit the LPM. We lift segmentation function $\Gamma$ to event logs, $\Gamma_{L,LPM}{=}\{\Gamma_{\sigma,LPM}|\sigma{\in}L\}$. An alignment-based \cite{Adriansyah2011} implementation of such a segmentation function $\Gamma$ is proposed in \cite{Tax2016}.\looseness=-1

We define $\mathit{activities}(a,L){=}|\{e{\in}\sigma|\sigma{\in} L{\land}\pi_\mathit{activity}(e){=}a\}|$ as a function that maps each activity in event log $L$ to its occurrence frequency. $\mathit{activities}(L)$ represents the multiset with the frequencies of all activities that occur in $L$. $\mathit{events}(L)=\{e{\in}\sigma|\sigma{\in}L\}$ is the set of events of $L$. Note that functions $\mathit{activities}$ and $\mathit{events}$ can also be applied to $\Gamma_{L,LPM}$, which is itself an event log.

\section{Local Process Model Constraints and Utility Functions}
\label{sec:utility_functions}
The discovery of Local Process Models (LPMs) can be steered towards the business needs of the process analyst by using a combination of \emph{constraints} and \emph{utility functions}. Constraints can for example be used to find fragments of process behavior that lead to a loan application getting declined, or to find fragments that only describe loan applications above \euro{15} K and never those below. 
Utility functions can be used to discover LPMs that give insight in which fragments of process behavior are associated with high financial costs or long time delays.

Constraints are requirements that the LPM has to satisfy, therefore, we define constraints as functions that result in 1 when the requirement holds and is 0 when it does not hold. In a general sense they are defined as a function $c:X\rightarrow\{0,1\}$ where $X$ is the \emph{scope} on which the function operates. We distinguish four different scopes on which $X$ can be defined: \emph{trace-level} (T), \emph{event-level} (E), \emph{activity-level} (A), and \emph{model-level} (M). The class diagram in Figure \ref{fig:scope} contains annotations which indicate which classes are included in each of the scopes.


Utility functions indicate to what degree an LPM is expected to be interesting and helpful to answer the business question of the process analyst at hand, and are defined as functions $f:X\rightarrow\mathbb{R}$. Like constraints, utility functions can be defined on the four scopes indicated in the class diagram of Figure \ref{fig:scope}.\looseness=-1

Multiple utility functions and constraints can be combined to form one composite function that describes the total utility of an LPM given log $L$. Given constraints $c_1,c_2,\dots,c_n$ and utility functions $f_1,f_2,\dots,f_k$, the composite utility $u$ is defined as:

\noindent$u(L,LPM)=\prod_{i=1}^{n}c_i(L,LPM)\cdot\sum_{j=1}^{k}f_j(L,LPM)$.

\noindent An LPM needs to satisfy all constraints $c_i$ in order to have a utility larger than zero, i.e., $\exists c_i(L,LPM){=}0\implies u(L,LPM){=}0$. Where the original LPM discovery method \cite{Tax2016} selects and ranks LPMs based on support, utility allows for goal-oriented selection and ranking, i.e., LPMs are ranked based on their utility. 

Utility functions and constrains of the trace, activity, and event-level scopes are all defined on a combination of an LPM and the event log. They define the utility of an LPM based on a function of log $L$ and its fragments that fit $LPM$, given by $\Gamma_{L,LPM}$. The trace, event, and activity-level scopes differ in the perspectives on $L$ and $\Gamma_{L,LPM}$ on which they operate. The model-level scope is defined solely on the LPM itself, and ignores the log argument $L$. Which concrete utility/constraint functions should be used depends on the business question of the process analyst. Figure \ref{fig:function_scopes} shows the arguments of utility and constraint functions on the four scopes. In the sections that follow we discuss definitions and properties of constraints and utility functions on each of these scopes and discuss several of their use cases.\looseness=-1

\tikzset{
	treenode/.style = {shape=rectangle, rounded corners,
		draw, align=center,
		top color=white},
	root/.style     = {treenode, font=\scriptsize},
	env/.style      = {treenode, font=\scriptsize},
}
\begin{figure}[b]
	\vspace{-0.35cm}
	\centering
	\scalebox{0.84}{
	\begin{tikzpicture}
	[
	grow                    = right,
	sibling distance        = 3.4em,
	level distance          = 13.5em,
	edge from parent/.style = {draw, -latex},
	every node/.style       = {font=\scriptsize},
	sloped,
	]
	\node [root,label=below:generic function] {$f(L,LPM)$\\
	$c(L,LPM)$}
	child { node [env,label=below:model-level] {$f^m(LPM)$\\
	$f^m(LPM)$}
		edge from parent node [below] {model-based} }
	child { node [env,label=below:trace-level] {$f^t(L,\Gamma_{L,LPM})$\\
	 $c^t(L,\Gamma_{L,LPM})$}
		child[] { node [env,label=below:activity-level] {$f^a(\mathit{activities}(L),\mathit{activities}(\Gamma_{L,LPM}))$\\
		 $c^a(\mathit{activities}(L),\mathit{activities}(\Gamma_{L,LPM}))$}
			edge from parent node [anchor=north east] {} }
		child[] { node [env,label=below:event-level] {$f^\epsilon(\mathit{events}(L),\mathit{events}(\Gamma_{L,LPM}))$\\
		 $c^\epsilon(\mathit{events}(L),\mathit{events}(\Gamma_{L,LPM}))$}
			edge from parent node [above] {} }
		edge from parent node [above] {Alignment-based} };
	\end{tikzpicture}}
	\caption{A tree of function specifications on different scopes.}
	\label{fig:function_scopes}
\end{figure}
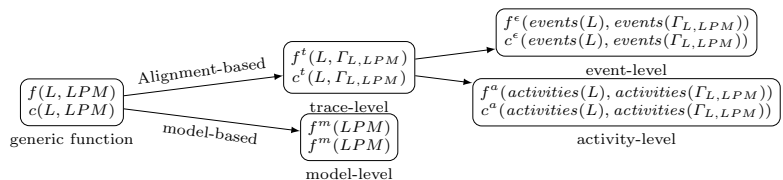

\subsection{Trace-level Constraints and Utility Functions}
\label{ssec:tl_mi_u}
Trace-level utility functions are the most general class of utility functions that are defined on the event log. Trace-level utility functions calculate the utility of an LPM by aggregating the utility over the trace-fragments that fit the LPM behavior and allow the utility of fitting trace fragment to depend on the events in the trace fragment, their event properties, and properties of the case itself. Trace level-utility can for example be used to discover LPMs that describe the events that explain a high share of the total financial cost associated to a case, or a high share of the total running time of a case.\looseness=-1

A trace-level utility function is a function $f^t(L,\Gamma_{L,\mathit{LPM}})$ that indicates the utility of the fitting trace fragments in $\Gamma_{L,\mathit{LPM}}$. Consider the LPM of Figure \ref{sfig:example_apn}, the example log consisting of one trace $\sigma$ in Figure \ref{sfig:example_trace}, and the segmentation of $\sigma$ on the LPM of Figure \ref{sfig:example_alignment}. Assume that $\sigma$ has a case property $\mathit{total\_cost}$ which indicates the total cost of the trace. An example of a trace-level utility function is $f_1^t(L,\Gamma_{L,LPM}){=}\sum_{\sigma'\in\Gamma_{L,LPM}}\sum_{e\in\sigma'}\frac{\pi_\mathit{cost}(e)}{\phi_{total\_cost}(\sigma')}$ which discovers LPMs that explain a large share of the total trace costs. Another example is a function $f_2^t(L,\Gamma_{L,LPM}){=}\sum_{\sigma'\in\Gamma_{L,LPM}}\frac{1}{\pi_\mathit{time}(\sigma'(|\sigma'|)){-}\pi_\mathit{time}(\sigma'(1))}$, which results in LPMs where the behavior described typically occurs in short time intervals. Trace-level constraints put requirements on the fitting trace segments $\Gamma_{L,\mathit{LPM}}$. All trace-level utility functions can be transformed into trace-level constraints by adding thresholds for a minimal or maximal value of the function.
\subsection{Event-level Constraints and Utility Functions}
\label{ssec:el_mi_u}
Event-level utility functions and constraints can be used when the business question of the process stakeholder concerns certain event properties, but does not concern the trace-context of those events. Such utility functions can e.g. be used to discover LPMs that describe process behavior fragments with high financial cost. Constraints on this level can e.g. be used to limit LPMs that solely describe events that are executed by certain resources, or, in certain time periods. The trace-level scope is more expressive than the event-level scope, allowing the analyst to formulate more complex utility functions and constraints, but at the same time, it makes it harder to formulate such functions compared to the more restricted event-level scope.\looseness=-1

An event-level utility function is a function $f^\epsilon(\mathit{events}(L),\mathit{events}(\Gamma_{L,\mathit{LPM}}))$ that indicates the utility of the fitting events in $\Gamma_{L,\mathit{LPM}}$. For the LPM, trace, and segmentation of Figure \ref{fig:example}, an example event-level utility function is $f_1^\epsilon(\mathit{events}(\mathit{L}),\\\mathit{events}(\mathit{\Gamma_{L,LPM}})){=}\sum_{e\in\mathit{events}(\mathit{\Gamma_{L,LPM}})}\pi_\mathit{cost}(e)$ which discovers LPMs with high costs. Note that $L$ itself is also in the domain, allowing us to formulate a utility function that optimizes the share of utility explained per activity $f_2^\epsilon(\mathit{events}(\mathit{L}),\\\mathit{events}(\mathit{\Gamma_{L,LPM}})){=}\sum_{a{\in}\Sigma_L}\frac{\sum_{e\in\mathit{events}(\mathit{\Gamma_{L,LPM}})}\pi_\mathit{cost}(e)\times(\pi_\mathit{activity}(e)=a)}{\sum_{e\in\mathit{events}(L)}\pi_\mathit{cost}(e)\times(\pi_\mathit{activity}(e)=a)}$.

An event level constraint is a function $c^\epsilon(\mathit{events}(L),\mathit{events}(\Gamma_{L,\mathit{LPM}}))$ and puts constraints on the events in $\mathit{events}(\Gamma_{L,\mathit{LPM}})$. An example is $c_1^\epsilon(\mathit{events}(\mathit{L}),\\ \mathit{events}(\mathit{\Gamma_{L,LPM}}))=f^\epsilon_1(\mathit{events}(\mathit{L}),\mathit{events}(\mathit{\Gamma_{L,LPM}})){\ge}500$, results in LPMs with a total value of at least \EUR{500}, or $c_2^\epsilon(\mathit{events}(\mathit{L}), \mathit{events}(\mathit{\Gamma_{L,LPM}}))=\forall_{e\in\Gamma_{L,LPM}}\pi_\mathit{cost}(e)\\{\ge100}$, which results in LPMs that never represent events with a value of less than \EUR{100}. Note that $c_2^\epsilon$ does not hold for our example trace and LPM, where the event with id 8 fits the LPM but only has value \EUR{50}. Therefore, this LPM will not be found by the LPM discovery technique when we use constraint $c_2^\epsilon$.


\subsection{Activity-level Constraints and Utility Functions}
\label{ssec:al_mi_u}
Activity-level utility functions and constraints define the utility of an LPM based on the frequency of occurrence of each activity in log $L$ and in $\Gamma_{L,LPM}$. Activity-level utility functions can for example be used by the process analyst to specify that he is more interested in some activities of high impact (e.g., lawsuits, security breaches, etc.) than in others, resulting in LPMs that describe the frequent behavior before and after such events. With activity-level constraints the process analyst can set a hard constraint on activity occurrences.
The stakeholder/analyst can specify a function $f_1^a(\mathit{activities}(L),\mathit{activities}(\Gamma_{L,\mathit{LPM}}))$, which indicates how interested he is in each activity. Note that a utility function with equal importance assigned to all activities, i.e., $f^a(\mathit{activities}(L),\mathit{activities}(\Gamma_{L,\mathit{LPM}}))=|\mathit{activities}(\Gamma_{L,\mathit{LPM}})|$, results in support-based LPM discovery as described in \cite{Tax2016}. 

Such utility functions can for example be used to get insight in the relations with other activities of some particular high-impact activities that the process analyst is interested in. Note that the occurrence of such activities in the log can be infrequent, in which case traditional LPM discovery without the use of utility functions and constraints is unlikely to return LPMs that concern those activities.\looseness=-1

Adding a transition representing a zero-utility activity to an LPM can never increase its total utility, therefore, activity-level utility functions that assign zero value to some activities speed up discovery by limiting the LPM search space $\mathit{LPMS}(L)$, because LPMs with zero-utility activities do not have to be evaluated.

\subsection{Model-level Utility Constraints and Utility Functions}
Model-level utility functions and constrains can be used when a process analyst has preference or requirements for specific structural properties of the LPM. They have the form $f^m(LPM)$ and $c^m(LPM)$, i.e., they are independent of the log and dependent only on the LPM itself. When a process analyst for example wants to analyze the behavior that leads to the execution of a certain activity $a$ which he is interested in, he can use a model-level constraint that enforces that all elements of $\mathcal{L}(LPM)$ end with $a$.

Generally we are interested in models that somehow represent the event log. Therefore, model-level utility functions and constraints are often not very useful on their own, but they become useful when combining them with utility functions and constraints on the event log, i.e., on the activity-level, event-level, or trace-level.\looseness=-1

$\Gamma_{L,\mathit{LPM}}$ does not need to be calculated to determine whether a model-level constraint is satisfied, because model-level constraints are defined solely on the model. Therefore, model-level constraints can also be used to speed up LPM discovery by limiting the search space of models $\mathit{LPMS}(L)$ to its subspace for which the model-level constraints hold.

\subsection{Composite Utility Functions}
\label{ssec:al_md_u}
Utility functions and constraints on the different levels can be combined into one single utility function. In the beginning of this section we defined the utility of a LPM for a given event log in the following way:

\noindent$u(L,LPM)=\prod_{i=1}^{n}c_i(L,LPM)\cdot\sum_{j=1}^{k}f_j(L,LPM)$

\noindent The individual constraints $c_i$ and the individual utility functions $f_i$ can be defined on any of the levels discussed in the previous sections to form one composite, possibly multi-level, utility function. The total utility $u$ is defined as an unweighted sum over the individual utility functions $f_i$. Note that this still allows the process analyst to give priority to one utility function over another, as weights can be included as a part of the utility function itself by multiplying the utility function with a constant.

The presented framework of utility functions and constraints generalizes the method described in \cite{Tax2016} and all LPM quality metrics presented there are instantiations of utility functions. An example is the \emph{support} metric, which is an activity-level utility function. The minimum threshold for support as proposed in \cite{Tax2016} is an example of an activity-level constraint. Another quality metric introduced in \cite{Tax2016} is \emph{determinism}, which is inversely proportional to the average number of enabled transitions in LPM during replay of the aligned event log. \emph{Determinism} is an example of a composite utility function, consisting of a model-level and a trace-level component.

Note that the trace-level and event-level utility functions are not limited to continuous-valued properties. Many event logs contain ordinal event properties, such as \emph{risk}, or, \emph{impact}, which can take values such as \emph{low}, \emph{medium}, or \emph{high}. Event-level utility functions can be applied to such event properties by specifying a mapping from the possible ordinal values to continuous values.

\section{Case Studies}
\label{sec:case_study_1}
In this section we describe two case studies on real life event logs. The first log originates from an IT service desk of a large Dutch financial institution. The second log originates from the traffic fine handling process by the Italian police.\looseness=-1

\subsection{IT Service Desk}
The IT service desk event log\footnote{http://dx.doi.org/10.4121/uuid:c3e5d162-0cfd-4bb0-bd82-af5268819c35} is an event log that was made publicly available as part of the \emph{Business Process Intelligence Challenge 2014}. The data set contains \emph{incident} events which represent disruptions of IT-services within a large financial institution. Each incident event is associated with one or more \emph{interactions}, which represent the calls and e-mails to the service desk agents that are related to this incident. When an incident occurs, it is assigned to an operator, who either solves the issue, or reassigns it to a colleague having more knowledge. For each incident event several properties are recorded, amongst others:
\begin{description}
	\item[Service Component WBS] This is a number that identifies the service component involved in the incident.
	\item[Configuration Item] This contains the type (i.e., laptop, server, software application, etc.) of the service component that the incident concerns. Each \emph{service component WBS} belongs to one \emph{configuration item}.
	\item[Impact] The impact of the service disruption to the customers as assessed by the operator. This property takes integer values from 1 to 5.
	\item[Closure code] A code which classifies the cause of the service disruption, e.g., user error, software error, hardware error.
	\item[CausedBy] Incidents of a service component have another service component as root cause of the service disruption. This field contains the \emph{service component WBS} number of the root cause service component.
	\item[Number of interactions] The number of calls to the IT service desk that are related to this incident.
	\item[Number of reassignments] The number of times that this incident was reassigned from one IT service desk operator to another.
\end{description}

We group together events by their \emph{service component WBS} number, creating one case per service component consisting of incidents that these service components are involved in. We set the activity label of each incident event to a combination of the \emph{closure code} and the \emph{causedBy} attribute separated by the pipe character ($|$). The resulting event log contains 313 traces, 25,262 events and 944 activities.\looseness=-1
\begin{figure}[t]
	\centering
	\includegraphics[width=\linewidth]{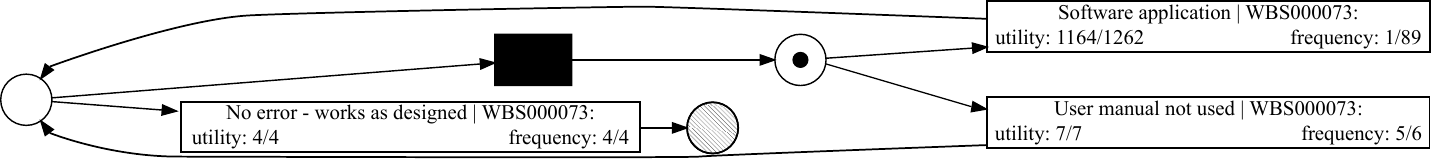}
	\smallbreak
	\includegraphics[width=\linewidth]{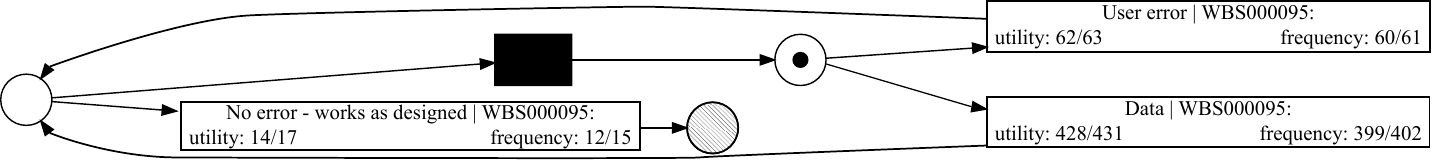}
	\caption{Two LPMs discovered from the BPI'14 event log when using event property \emph{number of interactions} as utility function.}
	\label{fig:bpic14_interactions}
	\vspace{-0.45cm}
\end{figure}

Assume now that we are a process analyst concerned with the business question: ``which process fragments are related to high numbers of e-mails and phone calls to the IT service desk?''. To answer this question we formulate utility function $f^\epsilon(\mathit{events}(L),\mathit{events}(\Gamma_{L,\mathit{LPM}}))=\sum_{e\in\mathit{events}(\Gamma_{L,\mathit{LPM}})}\pi_\mathit{number\_of\_interactions}(e)$. Figure \ref{fig:bpic14_interactions} shows the two LPMs with the highest utility that we discovered from the IT service desk event log using this utility function. 

The total utility of the top LPM of Figure \ref{fig:bpic14_interactions} is 1175, indicating that the behavior of this LPM explains 1175 calls and e-mails to the IT service desk. 
The LPM shows that in total 1262 calls and e-mails to the IT service desk were associated with incidents with closure code \emph{software application} that were caused by \emph{WBS000073}, 1164 of which were associated with such an incident that fits the LPM. Only one single incident with this closure code and causedBy number out of the total 89 in the log fit the LPM behavior, however, this single incident had high impact as it caused 1164 of the 1262 interactions. Finally, the LPM ends with an incident with closure code \emph{No error - works as designed}. This indicates that this software application contains a feature that is working properly according to the IT service desk (``works as designed''), while bank employees perceive it as a problem, resulting in a lot of traffic to the IT service desk. Because only one single \emph{software application $|$ WBS000073} incident fits the behavior of this LPM, one could say that it describes an anomaly rather than a pattern. Note that the discovery of such anomaly-type LPMs is a result of the way we defined our utility function, which allows for highly skewed distribution of utility over events.

The second LPM shows a similar pattern, but it concerns incidents caused by server based software application \emph{WBS000095}. This LPM has a total utility of 504, meaning that it describes in total 504 calls and e-mails to the IT service desk. The model shows that 60 out of 61 \emph{user errors} are eventually followed by an incident with closure code \emph{No error - works as designed}. These 60 incidents together generate 62 interactions with the IT service desk. Note that 399 data incidents related to this software application, causing 428 interactions with the service desk, also resulted in an incident with this closure code.

In the two LPMs we see that two service components (\emph{WBS000073} and \emph{WBS000095}) are the main cause of call and e-mails to the IT service desk. Furthermore, for both service components, the LPMs end in \emph{No error - works as designed} events, which could indicate that such problems could be prevented. \looseness=-1

\subsection{Traffic Fines}
The road traffic fine management event log\footnote{http://dx.doi.org/10.4121/uuid:270fd440-1057-4fb9-89a9-b699b47990f5} is an event log where each case refers to a traffic fine. Each case starts with a \emph{create fine} event, which has a property \emph{amount} that specifies the amount of the fine. \emph{Payment} events have a \emph{paymentAmount} property, which indicates how much has been paid. Payment of the total fine amount can be spread out over multiple payments. Some fines are paid directly to the police officer when the fine is given, and some are sent by mail, in which case there is a \emph{send fine} event. \emph{Send fine} events have a property \emph{expense}, which contains an additional administrative cost which adds to the total amount that has to be paid. When a fine is not paid in time, an \emph{add penalty} event occurs which has an \emph{amount} property that updates the fine amount set in the \emph{create fine} event. If a fine is still not paid after the added penalty it is \emph{send for credit collection}. Furthermore, a fine can be appealed at the prefecture and, at a later stage, can be appealed in court. In total the traffic fines event log contains 150,370 traces, 561,470 events, and 11 activities.

\begin{figure}[t]
	\renewcommand*\thesubfigure{\arabic{subfigure}} 
	\subfloat[total utility= \EUR{12.6} million]{
		\includegraphics[width=\linewidth]{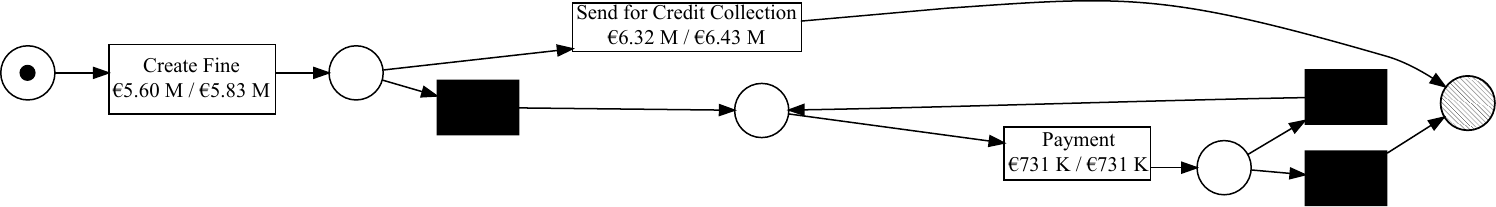}
	}\\
	\subfloat[total utility= \EUR{10.5} million]{
		\includegraphics[width=\linewidth]{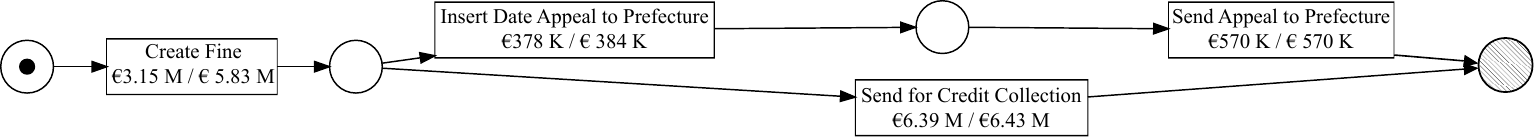}
	}\\
	\subfloat[total utility= \EUR{9.6} million]{
		\includegraphics[width=\linewidth]{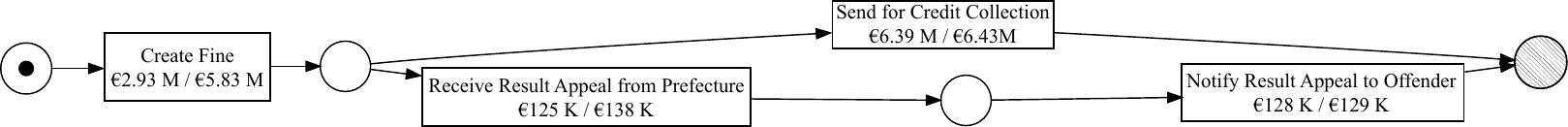}
	}
	\caption{The top three LPMs in terms of total utility as discovered from the traffic fine log, using \emph{remaining amount to pay} as utility.}
	\label{fig:road_fine_remaining_amont_utility}
	\vspace{-0.45cm}
\end{figure}

Assume we are a process analyst concerned with the business question: ``which process fragments describe fines where the remaining amount to be paid is high?''. Figure \ref{fig:road_fine_remaining_amont_utility} shows the top three LPMs that we discovered from the traffic fine log using a trace-level utility function that defines utility as the remaining amount that is still to be paid at the time of the event. This utility function has a trace-level scope, as it is calculated from the latest seen \emph{amount} in the case (either from a \emph{create fine} or \emph{add penalty} event) plus the \emph{expense} fee if there is a \emph{send fine} event minus the \emph{paymentAmount} values of all \emph{payment} events seen in the trace so far.\looseness=-1

The first LPM in Figure \ref{fig:road_fine_remaining_amont_utility} shows that in total \EUR{5.83} million has been created in fines, out of which \EUR{5.60} million was either send to credit collection or payments have been received for them. The remaining \EUR{200} thousand correspond to recent fines that have not yet been paid but are not yet due to be sent for credit collection. The total amount of payments received after a \emph{create fine} event is \EUR{731} thousand, which is surprisingly low comparable to the total of \EUR{5.83} million. Noteworthy is that fines representing a total value of \EUR{6.43} million are send to credit collection, which is even more than the fines representing a value of \EUR{5.83} million that were created in the first place. That the total value of fines at credit collection is higher than the total value of fines that were handed out is because of added penalties and, to a lesser degree, added expenses. Finally, fines representing a value of \EUR{6.32} million that were sent to credit collection out of the total \EUR{6.43} million value of fines that were sent to credit collection fit the control flow pattern of the LPM, i.e., they occur after a \emph{create fine} and are in an XOR-construct with \emph{payment} events. Since we know that all traces start with a \emph{create fine} event, the only possible explanation left is that fines representing a value of \EUR{6.32} million that were sent to credit collection have not received a single payment before being send to credit collection, while for the remaining fines representing a value of \EUR{110} thousand that were sent to credit collection at least one partial payment of the fine was already received.

The second LPM in Figure \ref{fig:road_fine_remaining_amont_utility} shows that fines representing a value of \EUR{3.15} out of the total value of fines of \EUR{5.83} million was either sent to credit collection or appealed at the prefecture. The appeal procedure starts with an \emph{insert data appeal to prefecture} event which is later followed by a \emph{send appeal to prefecture} event. The LPM shows that for the appealed fines, the total value is \EUR{378} thousand at the time of \emph{insert date appeal to the prefecture}, but added penalties for late paying raise the total appealed amount to \EUR{570} thousand at the time that the appeal is actually send to prefecture. This means appealed fines on average multiply 1.5x in value during the appeal procedure as a result of penalties being added for the payment term being overdue. Finally, the numbers in the \emph{send for credit collection} transition show that only a very small portion of the total value of fines that were sent to credit collection were followed by an appeal procedure.\looseness=-1

The third LPM in Figure \ref{fig:road_fine_remaining_amont_utility} shows a pattern similar to the second LPM, with the difference that it now describes two later steps in the appeal to prefecture procedure. At the \emph{receive result appeal from prefecture} step there is a total value of \EUR{138} thousand. This is considerably lower than the fines representing values of \EUR{378} thousand and \EUR{570} thousand respectively that we found for the earlier steps of the appeal procedure. This shows that there are appealed fines representing a value of \EUR{570} thousand - \EUR{138} thousand = \EUR{432} thousand which did not receive the appeal results. These appeals were either withdrawn before the verdicts on these appeals were made, or these appeals are still waiting for a verdict. Furthermore, fines representing a value of \EUR{125} thousand out of the total value of \EUR{138} thousand of fines at \emph{receive result appeal from prefecture} fit the control flow of the pattern, indicating that fines representing a value of \euro{13} thousand which received a result for appeal were either not (yet) notified, or they have been \emph{sent for credit collection} prior to the appeal.
\section{Conclusions \& Future Work}
\label{sec:conclusion}
This paper presents a framework of utility functions and constraints for Local Process Models (LPMs) that allows for combinations of utility functions and constraints on different scopes: on the activity, event, trace, and model level. We formalize utility functions on each of the levels and provide examples of how they can be used. Finally, we show on real-life event logs that the utility functions and constraints can be used to discover insightful LPMs that cannot be obtained using existing support-based LPM discovery.\looseness=-1

When a model e.g. allows for one execution of activity $a$ while multiple executions of $a$ are observed in the log, different alignments are possible depending on the choice of $a$ for the synchronous move and the ones for log moves. However, these events do not necessarily have equal utility. Thus, the utility of the LPM depends on the alignment  returned by the alignment algorithm. As future work, we plan to make alignments utility-aware, so that the optimal alignment leads to the highest LPM utility.

\bibliographystyle{splncs03}
\bibliography{bibliography}
\end{document}